\documentstyle[preprint,prc,aps]{revtex}

\begin{document}
\draft
\preprint{DE-FG06-88ER40427-04-N96}

\title{Quark-meson coupling model for finite nuclei}

\author{Peter G.\ Blunden$^{1,2}$
\thanks{On leave from Department of Physics, University of Manitoba,
Winnipeg, MB, Canada R3T~2N2}
and Gerald A.\ Miller$^2$}

\address{$^1$Institute for Nuclear Theory and $^2$Department of Physics
and Astronomy, Box 351560\\
University of Washington, Seattle, WA 98195-1560}
\date{\today}
\maketitle

\begin{abstract}
A Quark-Meson Coupling (QMC) model is extended to finite nuclei in the
relativistic mean-field or Hartree approximation. The ultra-relativistic
quarks are assumed to be bound in non-overlapping nucleon bags, and the
interaction between nucleons arises from a coupling of vector and scalar
meson fields to the quarks.  We develop a perturbative scheme for treating
the spatial nonuniformity of the meson fields over the volume of the nucleon
as well as the nucleus.  Results of calculations for spherical nuclei
are given, based on a fit to the equilibrium properties of nuclear matter. 
Several possible extensions of the model are also considered.
\end{abstract}


\section{Introduction}
Quantum Chromodynamics (QCD), a theory of quarks and gluons, is the
current paradigm for the strong interaction.  Nuclei are bound by the
strong interaction, but hadronic degrees of freedom account for the most
of the physical properties. This contrast suggests both questions and
opportunities.  One might ask how and why this standard picture of nuclear
physics as a system of clustered color-singlet objects emerges from the
fundamental theory.  The opportunities arise from the challenge of
discovering small (at normal nuclear densities) but interesting
corrections to the standard picture which arise from the underlying
constituents. One example was the discovery of the EMC
effect\cite{emcrev}, in which the structure function of a nucleon was
shown to be modified by the nuclear medium. 

Thus the need for a theory of nuclei that incorporates 
quark-gluon degrees of freedom, but also respects the 
vast body of information substantiating the standard picture,
is apparent. There are many possibilities, and here we 
wish to focus on the Quark-Meson Coupling (QMC) model.

In a recent series of papers Saito and Thomas\cite{st} have considered a
model for nuclear matter involving color-singlet clusters of quarks and
also mesons. This model is a variation on one originally proposed by
Guichon\cite{gu}, with corrections by Fleck et al.\cite{fleck}. The quarks
are assumed to be bound in non-overlapping nucleon bags, and the
interaction between nucleons arises from a coupling of meson fields to the
quarks.  The ultra-relativistic quarks are described by a mean-field Dirac
equation together with the MIT bag model boundary conditions\cite{MIT}.
The nucleons are assumed to be also described by the Dirac equation in the
effective mean fields arising from meson fields coupling to the quarks in
the nucleon. The model is conceptually similar to the QHD model of Walecka
and collaborators\cite{wal}, in which the meson fields couple to
point-like nucleons.  Indeed, in the limit that the quark mass becomes
very large the present model reduces to the Walecka model. But the
presence of the quark degrees of freedom provides important physical
content absent from QHD. 

The purpose of the present paper is to outline the formalism for applying
this model to finite nuclei.  Justifications for the adiabatic
approximation we use, in which the quarks do not respond to the motion of
the nucleon within the nucleus, have already been obtained by Guichon et
al.\cite{lastpaper}.  These authors also reported preliminary results for
$^{16}$O within a local density approximation to the nuclear matter
results. Here we provide a more detailed formalism which avoids the need
for some of the approximations they use. We also discuss extensions of the
model to account for other effects that may depend on the nuclear
density\cite{jenn}, and the use of a relativistic oscillator description
of the nucleon\cite{tegen} as an alternative to the bag model. 

An outline of the paper is as follows. The next section is concerned with
the formalism for both finite nuclei and nuclear matter.  Results are then
given for spherical nuclei, based on a fit of the model parameters to the
properties of nuclear matter.  Extensions of the model are then
considered, followed by a summary and outlook in the conclusions. 

\section{Theory of Finite Nuclei}
\subsection{Nuclear Energy and General Field Equations}

The nucleus is treated as a set of color singlet objects carrying
nucleonic quantum numbers. These objects (which we shall denote as bound
nucleons) are not the same as free nucleons because the forces that bind
the nucleus also influence the baryonic wave function. 

In the present model the forces between the bound nucleons are generated
by the exchange of vector ($\omega$) and scalar ($\sigma$) mesons coupling
to quarks. The static Hartree approximation is used to describe the meson
mean-fields and the bound nucleon single particle states. 

As in the QHD model\cite{wal}, we will assume the nucleons obey the Dirac
equation
\begin{equation}
\left[-\imath\bbox{\gamma}\cdot\bbox{\nabla}
+ M^*(\bbox{X})+\gamma^0 V_v(\bbox{X})\right]\psi_i(\bbox{X})=
E_i \gamma^0 \psi_i(\bbox{X}),
\label{diracn}
\end{equation}
with nucleon states labeled by quantum numbers $i$. The quantity
$M^*(\bbox{X})$ is the mass of the bound nucleon, which differs from the
mass of the free nucleon due to the influence of the scalar potential
generated by the sigma field on the quarks inside the nucleon. As will be
shown, $M^*(\bbox{X})$ is a nonlinear function of the sigma field in the
QMC models. $V_v(\bbox{X})$ is the time component of the vector potential,
and depends linearly on the time component of the omega field.  The vector
component of the omega field averages to zero in the nuclear rest frame,
so we drop it here. 

The total nuclear energy $\cal E$ is obtained as a sum of the nucleon and
meson field energies:
\begin{equation}
{\cal E}=\sum_{i=occ} E_i + E_v +E_s.
\label{etot}
\end{equation}
The first term is a sum over the occupied positive energy nucleon states.
The contribution of the vector mesons to the total energy is given by
\begin{equation}
E_v=-\case{1}{2}\int d^3r\; \omega
(\bbox{r})\left(m_v^2-\nabla^2\right) \omega(\bbox{r}),
\label{eomega}
\end{equation}
and the sigma meson contribution is
\begin{equation}
E_s=\case{1}{2}\int d^3r\;\sigma(\bbox{r})\left(m_s^2-\nabla^2\right)
\sigma(\bbox{r}).
\label{esigma}
\end{equation}
The mass of the vector meson is taken here to be the same as the omega
($m_v$=783 MeV), whereas the mass of the scalar meson is a parameter of
the model, typically in the range $m_s\sim 400-550$ MeV. 

The fields $\omega(\bbox{r})$ and $\sigma(\bbox{r})$ provide vector
\begin{equation}
V_v^q(\bbox{r})=g^q_v\;\omega(\bbox{r}),
\label{vqv}
\end{equation}
and scalar quark potentials
\begin{equation}
V_s^q(\bbox{r})=g^q_s\;\sigma(\bbox{r}),
\label{vqs}
\end{equation}
that act on confined quarks located at $\bbox{r}$.  Here $g_v^q$ and
$g_s^q$ are the relevant quark-meson coupling constants. The quark
potentials determine the quark wave functions and densities, which in turn
are needed to obtain the functions $\omega(\bbox{r})$ and
$\sigma(\bbox{r})$. Thus there is an additional self-consistency
requirement in the present model --- the quark and nuclear wave functions
must be determined together. 

The meson field equations are derived by minimizing $\cal E$ with respect
to variations in the meson fields $\omega(\bbox{r})$ and
$\sigma(\bbox{r})$, yielding
\begin{eqnarray}
\left(-\nabla^2 + m_v^2 \right) \omega(\bbox{r}) &=& 
\int d^3 X\;\rho_v(\bbox{X})
{\delta V_v(\bbox{X})\over \delta \omega(\bbox{r})},\\
\left(-\nabla^2 + m_s^2 \right) \sigma (\bbox{r}) &=& -
\int d^3 X\;\rho_s(\bbox{X})
{\delta M^*(\bbox{X})\over \delta \sigma(\bbox{r})}.
\label{lapsgen}
\end{eqnarray}
Here $\rho_{v,s}(\bbox{X})$ represents the vector ($v$) and scalar ($s$)
nucleon densities at a position $\bbox{X}$, with
\begin{eqnarray}
\rho_v(\bbox{X})&=&\sum_{i=occ}\psi_i^\dagger
(\bbox{X})\psi_i(\bbox{X}),\\
\rho_s(\bbox{X})&=&\sum_{i=occ}{\bar\psi_i}(\bbox{X})\psi_i(\bbox{X}).
\end{eqnarray}

The relation between the nuclear potentials and the meson fields will be
derived in the next sub-section. In anticipation of these results, we set
\begin{equation}
{\delta V_v(\bbox{X})\over \delta \omega(\bbox{r})} =
3 g_v^q \rho_v^q(\bbox{r}-\bbox{X};\bbox{X}),
\end{equation}
and, for the simplest of the QMC models,
\begin{equation}
{\delta M^*(\bbox{X})\over \delta \sigma(\bbox{r})} =
-3 g_s^q \rho_s^q(\bbox{r}-\bbox{X};\bbox{X}).
\label{dmsigma}
\end{equation}
The quantities $\rho_v^q(\bbox{r}-\bbox{X};\bbox{X})$ and
$\rho_s^q(\bbox{r}-\bbox{X};\bbox{X})$ represent the nucleonic expectation
values of the vector and scalar quark field operators for a nucleon with
center-of-mass located at $\bbox{X}$. The variable $\bbox{r}-\bbox{X}$ is
the displacement between a quark and the center of its nucleon. For
simplicity, we have assumed that the three quarks inside the nucleon have
identical masses, and therefore identical density distributions. The
calculation of these quark densities is described in the next sub-section. 

With these results, the meson field equations become:
\begin{eqnarray}
\left(-\nabla^2 + m_v^2 \right) \omega(\bbox{r}) &=& 3 g_v^q
\int d^3 X\;\rho_v(\bbox{X}) \rho_v^q(\bbox{r}-\bbox{X};\bbox{X}),
\label{lapv}\\
\left(-\nabla^2 + m_s^2 \right) \sigma (\bbox{r}) &=& 3 g_s^q
\int d^3 X\;\rho_s(\bbox{X}) \rho_s^q(\bbox{r}-\bbox{X};\bbox{X}).
\label{laps}
\end{eqnarray}
Thus the source term for $\omega(\bbox{r})$ in Eq.\ (\ref{lapv}) is a
convolution of the vector density of the nucleon with the vector density
of the quarks in the nucleon. Similarly, the source term for
$\sigma(\bbox{r})$ in Eq.\ (\ref{laps}) is a convolution of the scalar
density of the nucleon with the scalar density of the quarks in the
nucleon. This distinction, which was not present in the original work of
Guichon\cite{gu}, was first pointed out by Fleck et al.\cite{fleck}, who
derived it from an analysis of a boost of the composite system. In
Ref.\cite{lastpaper} the convolution of densities has been replaced by an
evaluation of the nuclear densities at the point $\bbox{r}$.  This is
equivalent to a local density approximation to the nuclear matter result,
where the nucleon densities are constant over the volume of the nucleon. 

Strictly speaking, Eq.\ (\ref{lapv}) is incomplete.  The $\omega$-meson
couples to a quark baryon current of the simple form $j_q^\mu=\bar\psi_q
\gamma^\mu \psi_q$.  However, the nucleonic matrix element of $j_q^\mu$,
which is the effective baryon current of the composite nucleon ($\psi$),
will have both a $\bar\psi \gamma^\mu\psi$ term and an ``anomalous'' term
$\partial_\nu\bar\psi \sigma^{\mu\nu}\psi$.  This is the well-known
``tensor'' coupling for vector mesons and photons.  It is proportional to
the anomalous magnetic moment of the nucleon, and emerges naturally in the
present model.  In the mean-field approximation there is a non-vanishing
contribution from this term for the time-like part of the current
($\mu=0$).  We give a more detailed discussion of this point in Appendix
A.  Because the tensor coupling of the $\omega$-meson is known to be
small, we choose to ignore its influence in the computations performed
here. 
 
The essential feature of the present sub-section is that the
self-consistent solution of Eqs.\ (\ref{diracn}), (\ref{lapv}) and
(\ref{laps}) give the energy and wave function of the nucleus. These
equations are similar to the ones of QHD\cite{wal}, but with the essential
difference being the dependence on the quark vector and scalar densities. 

\subsection{Quark Wave Functions and Densities}

Computing the quark wave functions and the resulting nucleonic vector
potential $V_v(\bbox{X})$ and effective mass $M^*(\bbox{X})$ starts with
using a fairly general representation of the field equation of a confined
quark,
\begin{equation}
\left[-\imath \gamma^0{\partial\over \partial t}
-\imath\bbox{\gamma}\cdot\bbox{\nabla}_{\bbox{r}} + m_q
+V_{con}(\bbox{r}-\bbox{X})\right]\;\psi_q^f(t,\bbox{r}-\bbox{X})= 0.
\label{fdiracq}
\end{equation}
Here $\bbox{X}$ is the position of the center of the free nucleon, $m_q$
is the bare quark mass, and the confining potential is defined as
$V_{con}(\bbox{r}-\bbox{X})$. This potential depends on the distance
between the quark and the center of its nucleon.  This central confining
potential can be understood to represent the MIT bag, when complemented by
the bag boundary conditions. In relativistic potential models one can take
$V_{con}(\bbox{r})= C r^n$, as did Tegen et al.\cite{tegen}, who consider
$n=2$ and $n=3$. $V_{con}$ can also be obtained as a Hartree approximation
to a more realistic treatment of confinement based on two or three quark
interactions.  The superscript $f$ is meant to denote that the nucleon is
free. 

Next we consider how Eq.\ (\ref{fdiracq}) is modified when the nucleon is
bound. Each quark feels a vector $V_v^q(\bbox{r})$ and scalar
$V_s^q(\bbox{r})$ potential. If the nucleon is centered at a position
$\bbox{X}$, then one expects that the quark wave functions will depend on
both $\bbox{r}$ and $\bbox{X}$. Thus we obtain
\begin{equation}
\left[-\imath \gamma^0{\partial\over \partial t}
-\imath\bbox{\gamma}\cdot\bbox{\nabla}_{\bbox{r}} + m_q - 
V_s^q(\bbox{r}) +V_{con}(\bbox{r}-\bbox{X})
+ \gamma^0 V_v^q(\bbox{r})\right]\;\psi_q(t,\bbox{r}-\bbox{X},\bbox{r})= 0.
\label{diracq}
\end{equation}
This dependence can be fairly complicated. The $\omega$ and $\sigma$
fields and resulting quark potentials are functions of $\bbox{r}$, so that
such fields are not functions of $\bbox{r}-\bbox{X}$, and are not central
in the frame of the nucleon.  Hence the ground state of a bound nucleon
will not be spherically symmetric in general. 

To handle the difficulties that this entails, we develop a new
approximation scheme based on the notion that
\begin{equation}
V_{v,s}^q(\bbox{r})\;\psi_q(t, \bbox{r}-\bbox{X},\bbox{r})
\approx U_{v,s}(\bbox{X})\;\phi_q(t, \bbox{r}-\bbox{X};\bbox{X}),
\label{notion}
\end{equation}
where $U_{v,s}(\bbox{X})$ is some suitable averaging of
$V_{v,s}^q(\bbox{r})$ over the volume of the nucleon. The notation
$\phi_q(t, \bbox{r}-\bbox{X};\bbox{X})$ is meant to denote that the quark
wave function depends on the coordinate variable $\bbox{r}-\bbox{X}$, and
is only an implicit function of $\bbox{X}$ through $V_{v,s}(\bbox{X})$.
This notion is plausible because $\psi_q(t,\bbox{r}-\bbox{X},\bbox{r})$
vanishes if $\bbox{r}-\bbox{X}$ is much bigger than the mean radius of a
nucleon, and $V_s^q(\bbox{r})$ and $V_v^q(\bbox{r})$ are not expected to
vary much over this region. The variation of the nuclear fields is
governed by two scales --- the nuclear radius, and the nuclear skin
thickness (distance at the surface over which the density decreases from
90\% to 10\% of its maximum value) of about 2.5 fm.  Either distance is
much larger than the radius of the nucleon, so that replacing $V_s^q$ and
$V_v^q$ by some suitable average over the volume of the nucleon seems
reasonable, at least at the start. 

The procedure (but not the theory) of Ref.\cite{lastpaper} is to evaluate
the external fields at the nucleon center and neglect the variation of
these fields within the volume of a single nucleon.  This means that their
calculations replace the approximate sign in Eq.\ (\ref{notion}) by an
equality with $U_{v,s}(\bbox{X})=V_{v,s}^q(\bbox{X})$. 

Here we intend to develop a more complete formalism, based on the
plausibility of Eq.\ (\ref{notion}). We define an ``average'' potential
$U_{v,s}(\bbox{X})$ and a residual interaction
$\Delta U_{v,s}(\bbox{r},\bbox{X})$ such that
\begin{equation}
V_{v,s}^q(\bbox{r})
= U_{v,s}(\bbox{X}) + \Delta U_{v,s}(\bbox{r},\bbox{X}).
\label{udef}
\end{equation}
This equation is clearly a tautology. However, we may choose 
the potentials $U_{v,s}(\bbox{X})$ so as to make the first-order 
perturbation theory evaluation of $\Delta U_{v,s}(\bbox{r},\bbox{X})$ 
vanish. This suggests that we can develop a convergent perturbative
treatment of $\Delta U_{v,s}$.

The feature that $U_{v,s}$ depends only on $\bbox{X}$ allows us to
simplify the solution of the Dirac equation (\ref{diracq}).  We take
\begin{equation}
\psi_q(t,\bbox{r}-\bbox{X},\bbox{r})=\sum_n\;c_n\;
e^{-i \left[\epsilon^*_n(\bbox{X}) + U_v(\bbox{X})\right]t}\; 
\phi_n(\bbox{r}-\bbox{X};\bbox{X}),
\end{equation}
where the quark energies $\epsilon^*_n(\bbox{X}) $ and wave functions
$\phi_n(\bbox{r}-\bbox{X};\bbox{X})$ of quarks with quantum
numbers $n$ are determined by solving the equation
\begin{equation}
\left[-\imath\bbox{\gamma}\cdot\bbox{\nabla}_{\bbox{r}} + m_q - 
U_s(\bbox{X})
+V_{con}(\bbox{r}-\bbox{X})\right]\;\phi_n(\bbox{r}-\bbox{X};\bbox{X})
= \epsilon^*_n(\bbox{X})\;\gamma^0 \phi_n(\bbox{r}-\bbox{X};\bbox{X}).
\label{diracb}
\end{equation}
The quantity $c_n$ is to be determined by perturbative calculations with
the residual interaction $\Delta U_{v,s}(\bbox{r},\bbox{X})$. 

Comparing Eq.\ (\ref{diracb}) with the equation for the free nucleon
(\ref{fdiracq}), it is clear that the quark wave function $\phi$ in the
medium is of the same form as the free quark wave function $\phi^f$, but
with the bare quark mass $m_q$ replaced by an effective mass
\begin{equation}
m_q^*(\bbox{X}) \equiv m_q- U_s(\bbox{X}).
\end{equation}
Thus the quark energies and densities (or wave functions) are implicitly
functions of $\bbox{X}$ only via $m_q^*(\bbox{X})$.  This is properly
expressed by the notation $\epsilon^*(m_q^*(\bbox{X}))$ and
$\rho_{v,s}^q(\bbox{r}-\bbox{X};m_q^*(\bbox{X}))$.  We will continue to
use the simpler notation $\epsilon^*(\bbox{X})$ and
$\rho_{v,s}^q(\bbox{r}-\bbox{X};\bbox{X})$ with this understanding. We
emphasize that the vector interactions have no effect on the nucleon
properties in the medium other than an overall phase in the wave function,
which results in a shift in the nucleon energies. 

As a first step let us assume that the nucleon consists of three quarks
each in a state with $\kappa=-1$.  This state is treated in perturbation
theory, so that state is simply $\phi_0$, which we shall simply denote as
$\phi$, i.e. $\phi\equiv \phi_0$.  Similarly $\epsilon\equiv \epsilon_0$. 
(In this case $c_n=c_n^{(0)}=\delta_{n,0}.$) These ground state quark wave
functions then determine the quark vector and scalar densities
\begin{eqnarray}
\rho_v^q(\bbox{\tilde r};\bbox{X})&=&\phi^\dagger(\bbox{\tilde r};\bbox{X})
\phi(\bbox{\tilde r};\bbox{X}),\label{rhovqdef}\\
\rho_s^q(\bbox{\tilde r};\bbox{X})&=&
\bar\phi(\bbox{\tilde r};\bbox{X})\phi(\bbox{\tilde r};\bbox{X}),
\label{rhosqdef}
\end{eqnarray}
subject to the normalization condition
\begin{equation}
\int d^3 \tilde r\; \rho_v^q(\bbox{\tilde r};\bbox{X}) = 1.\label{normv}
\end{equation}
We also introduce for convenience the definition
\begin{equation}
\int d^3 \tilde r\; \rho_s^q(\bbox{\tilde r};\bbox{X}) \equiv S(\bbox{X}).
\label{norms}
\end{equation}

$U_{v,s}(\bbox{X})$ is chosen according to the criterion that the
first-order perturbation theory evaluation of $\Delta
U_{v,s}(\bbox{r},\bbox{X})$ vanishes.  This is achieved with the
definitions
\begin{eqnarray}
U_v(\bbox{X})&\equiv&\int d^3r\;\rho_v^q(\bbox{r}-\bbox{X};
\bbox{X}) V^q_v(\bbox{r}),\label{uv}\\
U_s(\bbox{X})&\equiv&\int d^3r\;\rho_s^q(\bbox{r}-\bbox{X};
\bbox{X}) V^q_s(\bbox{r})\;/\;S(\bbox{X}).\label{us}
\end{eqnarray}
Explicitly then, the first-order corrections to the potential
$U_v(\bbox{X})$ and the quark energy $\epsilon^*(\bbox{X})$ are
\begin{eqnarray}
U_v^{(1)}(\bbox{X})&\equiv&\int d^3r\;\phi^\dagger(\bbox{r}-\bbox{X};
\bbox{X})\;\Delta U_v(\bbox{r},\bbox{X})\;\phi(\bbox{r}-\bbox{X};
\bbox{X})=0,\\
{\epsilon^*}^{(1)}(\bbox{X})&\equiv&\int d^3r\;
\bar\phi(\bbox{r}-\bbox{X};\bbox{X})\;
\Delta U_s(\bbox{r},\bbox{X})\;\phi(\bbox{r}-\bbox{X};\bbox{X})=0.
\end{eqnarray}
The quantity $\Delta U_{v,s}$ does cause a first-order change in the quark
wave function and a second order change in the potentials.  This is
discussed more fully in Appendix B. 

The net result is that Eqs.\ (\ref{uv}) and (\ref{us}) determine the mean
fields that act on the quarks. Our mean fields are computed by taking the
average over the relevant quark densities, whereas in Ref.\cite{lastpaper}
these fields are evaluated at the center of the nucleon. This difference
has a modest effect on the properties of finite nuclei, as discussed in
the results described in the next section. 

We may now obtain the nuclear vector and scalar potentials to be used in
Eq.\ (\ref{diracn}).  The term $U_v$ is present for each quark. Hence the
nuclear vector potential is three times the shift in the quark energy
caused by $U_v(\bbox{X})$, i.e.
\begin{equation}
V_v(\bbox{X})=3\;U_v(\bbox{X}).
\label{vector}
\end{equation}

The form of the effective mass $M^*(\bbox{X})$ depends on whether one uses
the relativistic potential models or the MIT bag model. In the
relativistic potential models we expect the mass of the nucleon to be just
the sum of the energies of the three quarks.  Hence, in the absence of any
correction for center-of-mass effects,
\begin{equation}
M^*(\bbox{X}) = 3\;\epsilon^*(m_q^*(\bbox{X})),
\end{equation}
where $\epsilon^*(m_q^*)$ is the energy eigenvalue of Eq.\ (\ref{diracb}).
It is convenient to introduce a nuclear scalar potential $V_s(\bbox{X})$
to be used in Eq.\ (\ref{diracn}), with
\begin{equation}
V_s(\bbox{X})\equiv M-M^*(\bbox{X}) =
3\;\left(\epsilon(m_q)-\epsilon^*(m_q^*)\right).
\label{scalar}
\end{equation}
This facilitates a comparison of QMC models with QHD\cite{wal}, and also
allows us to use the experimental proton and neutron masses in Eq.\
(\ref{diracn}) with the same scalar potential. 

In the MIT bag model with quarks of effective mass $m_q^*$, the ground
state solution to Eq.\ (\ref{diracb}) is\cite{st}
\begin{mathletters}
\begin{equation}
\phi(\bbox{r}) = {\cal N}_q \left(\begin{array}{c}
j_0(x_q r/R)\\ \imath\bbox{\sigma}\cdot\bbox{\hat r}
\beta_q j_1(x_q r/R)\end{array}\right)
{\chi_q \over (4 \pi)^{1/2}},
\end{equation}
where
\begin{eqnarray}
\epsilon^* &=& \Omega_q^* /R,\\
\Omega_q^* &=& \left[x_q^2 + (R m_q^*)^2\right]^{1/2},\\
{\cal N}_q^{-2} &=& 2 R^3 j_0^2(x_q)
\left[\Omega_q^* (\Omega_q^* -1) + R m_q^*/2\right]/x_q^2,\\
\beta_q &=& \left[(\Omega_q^* - R m_q^*)/(\Omega_q^* + R m_q^*)\right]^{1/2},
\end{eqnarray}
\end{mathletters}
and $\chi_q$ is the quark spinor. The eigenvalue $x_q$ is determined by
satisfying the linear boundary condition $j_0(x_q) = \beta_q j_1(x_q)$ at
the bag surface. 

We take the nucleon mass to be
\begin{equation}
M^*(\bbox{X}) = {3 \Omega_q^*-z \over R} + \case{4}{3} \pi R^3 B.
\label{mstar}
\end{equation}
Here $B$ is the bag constant and $z$ is a free parameter that is supposed
to account for zero-point motion, vacuum corrections, etc.  Our definition
is different than Fleck et al.\cite{fleck} and the initial work of Saito
and Thomas\cite{st}, who include a correction for spurious center-of-mass
motion.  The parameters $B$ and $z$ are determined by the free nucleon
mass $M$ for a given bag radius $R_0$, and are given in Table
\ref{table1}.  A possible medium-dependence of $B$ and $z$ is considered
in section \ref{ext}. 

Following the Born-Oppenheimer approximation, the bag is assumed to
respond instantaneously to changes in the nuclear environment.  Hence the
equilibrium condition,
\begin{equation}
\left.{\partial M^*\over \partial R}\right|_{R=R^*} = 0,
\end{equation}
also applies in the medium\cite{st}. In practice, for the models
considered in this paper, $R^*$ at nuclear matter densities is not very
different from the radius $R_0$ of a free nucleon bag.  Thus the dominant
contribution to $V_s(\bbox{X})$ is given by the single particle energies,
consistent with Eq.\ (\ref{scalar}). 

The desired results of this section are the vector and scalar potentials
that act on a nucleon, contained in Eqs.\ (\ref{vector}) and
(\ref{scalar}). These in turn depend on the mean vector and scalar quark
potentials, given in Eqs.\ (\ref{uv}) and (\ref{us}). The next step is to
obtain more specific forms for these potentials for use in our
calculations.  This is done in the next section. 

\subsection{Specific Evaluation of Nucleonic Potentials}\label{secspecific}

We wish to obtain the external vector and scalar mean-fields which act on
quarks in a nucleon centered at $\bbox{X}$. We shall first consider the
vector potential $V_v(\bbox{r})$.  The formal solution to Eq.\
(\ref{lapv}) is
\begin{equation}
V_v^q(\bbox{r}) = 3 (g_v^q)^2 \int d^3 X'\; \rho_v(\bbox{X}')
\int d^3 r'\; {e^{-m_v | {\bbox{r}}-{\bbox{r}}'|}
\over 4 \pi |\bbox{r}-\bbox{r}'|}\; \rho_v^q(\bbox{r}'-\bbox{X}';\bbox{X}').
\label{vvq}
\end{equation}
This is easily evaluated in momentum space.  Using the relation
\begin{equation}
{e^{-m_v | \bbox{r}-\bbox{r}'|}
\over 4 \pi |\bbox{r}-\bbox{r}'|} = \int {d^3 q\over (2 \pi)^3}\;
{e^{\imath \bbox{q}\cdot (\bbox{r}-\bbox{r}')}\over q^2+m_v^2},
\end{equation}
and changing variables to $\bbox{\tilde r} = \bbox{r}'-\bbox{X}'$,
we can rewrite Eq.\ (\ref{vvq}) as
\begin{equation}
V_v^q(\bbox{r}) = 3 (g_v^q)^2 \int {d^3 q\over (2 \pi)^3}\;
{e^{\imath {\bbox{q}}\cdot {\bbox{r}}}\over
q^2+m_v^2} \int d^3 X'\; e^{-\imath {\bbox{q}}\cdot {\bbox{X}'}}
\rho_v(\bbox{X}) \int d^3\tilde r\; e^{-\imath \bbox{q}\cdot
\bbox{\tilde r}}\rho_v^q(\bbox{\tilde r};\bbox{X}').
\end{equation}
The last integral above suggests that one may define a vector form factor
for bound nucleons: 
\begin{equation}
v(\bbox{q};\bbox{X})\equiv 
\int d^3\tilde r\; e^{-\imath \bbox{q}\cdot \bbox{\tilde r}}
\rho_v^q(\bbox{\tilde r};\bbox{X}).
\label{vform}
\end{equation}
Note that the normalization condition (\ref{normv}) ensures that
$v(0;\bbox{X})=1$. The use of the definition (\ref{vform}) leads to the
result
\begin{equation}
V_v^q(\bbox{r}) = 3 (g_v^q)^2 \int {d^3 q\over (2 \pi)^3}\;
{e^{\imath \bbox{q}\cdot \bbox{r}}\over
q^2+m_v^2} \int d^3 X'\; e^{-\imath \bbox{q}\cdot \bbox{X}'}
\rho_v(\bbox{X}') v(\bbox{q};\bbox{X}').\label{vvq2}
\end{equation}

The external vector potential felt by a nucleon centered at $\bbox{X}$ is
given according to Eqs.\ (\ref{uv}) and (\ref{vector}) by the convolution
\begin{eqnarray}
V_v(\bbox{X}) &=& 3 \int d^3 r\; \rho_v^q(\bbox{r}-\bbox{X};\bbox{X})
V_v^q (\bbox{r})\nonumber\\
&=& (3g_v^q)^2 \int {d^3 q\over (2 \pi)^3}\;
{e^{\imath \bbox{q}\cdot \bbox{X}}\over
q^2+m_v^2}\; v(\bbox{q};\bbox{X})
\int d^3 X'\; e^{-\imath \bbox{q}\cdot \bbox{X}'}
\rho_v(\bbox{X}') v(\bbox{q};\bbox{X}').
\label{vvn}
\end{eqnarray}
This quantity is easily recognized as the mean-field vector potential for
nucleons interacting via meson exchange with (position-dependent) form
factors. For the vector meson, the form factors are all that distinguishes
the present model from the QHD model\cite{wal,hs}. 

The quark scalar potential $V_s^q(\bbox{r})$ follows analogously from
Eq.\ (\ref{vvq2}).  We find
\begin{equation}
V_s^q(\bbox{r}) = 3 (g_s^q)^2 \int {d^3 q\over (2 \pi)^3}\;
{e^{\imath \bbox{q}\cdot \bbox{X}}\over
q^2+m_s^2} \int d^3 X'\; e^{-\imath \bbox{q}\cdot \bbox{X'}}
\rho_s(\bbox{X'}) s(\bbox{q};\bbox{X'}),
\label{vsq2}
\end{equation}
where the scalar form factor $s(\bbox{q};\bbox{X})$ for bound
nucleons is defined by the equation
\begin{equation}
s(\bbox{q};\bbox{X})\equiv 
\int d^3\tilde r\; e^{-\imath \bbox{q}\cdot \bbox{\tilde r}}
\rho_s^q(\bbox{\tilde r};\bbox{X}).
\label{sform}
\end{equation}
Unlike the quark vector form factor, there is no normalization constraint
on $s(0;\bbox{X})$, which is just the quantity $S(\bbox{X})$ of
Eq.\ (\ref{norms}). 

In analogy to Eq.\ (\ref{vvn}), the mean quark scalar potential
of Eq.\ (\ref{us}) is given by the convolution
\begin{eqnarray}
U_s(\bbox{X}) &=& \int d^3 r\; \rho_s^q(\bbox{r}-\bbox{X};\bbox{X})
V_s^q(\bbox{r})\;/\;S(\bbox{X})\nonumber\\
&=& {3(g_s^q)^2\over S(\bbox{X})} \int {d^3 q\over (2 \pi)^3}\;
{e^{\imath \bbox{q}\cdot \bbox{X}}\over
q^2+m_s^2}\; s(\bbox{q};\bbox{X})
\int d^3 X'\; e^{-\imath \bbox{q}\cdot \bbox{X}'}
\rho_s(\bbox{X}') s(\bbox{q};\bbox{X}').
\label{vvs}
\end{eqnarray}
This is the quantity used in the Dirac equation (\ref{diracb}) that
determines the quark energies $\epsilon^*$, which in turn determines the
external nuclear scalar potential $V_s(\bbox{X})$. 

The total energy $\cal E$ of Eq.\ (\ref{etot}) can now be given in terms
of the quantities we have just determined. Using Eqs.\ (\ref{eomega}),
(\ref{lapv}), (\ref{vqv}), (\ref{vvq2}) and (\ref{vvn}), we find for the
vector meson contribution to the total energy
\begin{equation}
E_v = - \case{1}{2} \int d^3 X\;\rho_v(\bbox{X}) V_v(\bbox{X}).
\label{eomega2}
\end{equation}
This is the same as in QHD, and follows from the linear dependence of
$V_v(\bbox{X})$ on the omega field $\omega(\bbox{r})$.  The effect is to
remove half the vector potential energy contribution contained in the
nucleon energies $E_i$ (see Eq.\ (\ref{diracn})).

For the scalar meson contribution, we use Eqs.\ (\ref{esigma}),
(\ref{laps}), (\ref{vqs}), (\ref{vsq2}) and (\ref{vvs}) to find
\begin{equation}
E_s = \case{1}{2} \int d^3 X\;\rho_s(\bbox{X})\;3 U_s(\bbox{X}) S(\bbox{X}).
\end{equation}
Because $V_s(\bbox{X})\neq 3 U_s(\bbox{X}) S(\bbox{X})$, this contribution
does not remove half the scalar potential energy contribution contained in
the nucleon energies.  However, this relation is approximately correct,
and becomes exact in the limit of large quark mass $m_q$.  In this limit,
$S(\bbox{X})\rightarrow 1$, and $V_s(\bbox{X})$ becomes a linear function
of $\sigma(\bbox{r})$.  One therefore expects to recover the results of
the QHD model. 

Finally, we limit our considerations to spherical nuclei, so that both the
quark and nucleon densities are spherically symmetric. The equations we
need are then: 
\begin{eqnarray}
V_v(X) &=& g_v^2 {2 \over \pi} \int_0^\infty
dq\; {q^2\over q^2+m_v^2}\;
j_0(q X) v(q;X) \int_0^\infty dX'\; X'^2\; j_0(q X') \rho_v(X')
v(q;X'),\label{vvnsph}\\
U_s(X) &=& {g_s^2 \over 3 S(X)} {2 \over \pi}
\int_0^\infty dq\; {q^2\over q^2+m_s^2}\; j_0(q X) s(q;X)
\int_0^\infty dX'\; X'^2\; j_0(q X') \rho_s(X') s(q;X'),\\
v(q;X) &=& 4\pi \int_0^\infty dr\; r^2\; j_0(q r) \rho_v^q(r;X),\\
s(q;X) &=& 4\pi \int_0^\infty dr\; r^2\; j_0(q r) \rho_s^q(r;X),
\end{eqnarray}
together with the Dirac equations for the quarks (\ref{diracb}) and for
the nucleons (\ref{diracn}). We have introduced the meson-nucleon coupling
constants $g_v\equiv 3g_v^q$ and $g_s\equiv 3g_s^q$. 

\subsection{Nuclear Matter Limit}

In the spirit of the usual approach to the QHD model\cite{wal}, we use the
equilibrium properties of isospin symmetric nuclear matter to fix the
parameters of the model. If we consider the limit in which the nucleus is
very large, the nucleon densities $\rho_v(X)$ and $\rho_s(X)$ can be
treated as constants ($\rho_B$ and $\rho_s$), as can the vector and scalar
fields.  Hence $U_v=V_v^q$ and $U_s=V_s^q$, so that $m_q^*=m_q-V_s^q$. Our
formalism will yield the nuclear matter results of Refs.\cite{st,fleck},
except for the correction due to the spurious motion of the
center-of-mass.  According to Ref.\cite{lastpaper}, this correction should
not be included. 

The composite nucleon obeys the Dirac equation with an effective mass
$M^*$, determined by Eq.\ (\ref{mstar}), and a vector potential $V_v = 3
V_v^q$.  Hence the nucleons have energy eigenvalues
\begin{equation}
E(k) = (k^2 + {M^*}^2)^{1/2} + 3 V_v^q.
\end{equation}
The total energy $\cal E$ is the sum of the energies of all nucleons below
the Fermi energy together with the contribution from the meson fields,
\begin{equation}
{\cal E} = 4 \int_0^{k_F} {d^3 k\over (2 \pi)^3}\;
\left[(k^2 + {M^*}^2)^{1/2} + 3 V_v^q\right] +
\case{1}{2}m_s^2 \sigma^2 - \case{1}{2} m_v^2 \omega^2.
\end{equation}
Minimizing $\cal E$ with respect to $\omega$ and $\sigma$ yields the
meson field equations (cf.\ Eqs.\ (\ref{lapv}) and (\ref{laps}))
\begin{eqnarray}
V_v^q &=& g_v^q\; \omega = {3 {g_v^q}^2\over m_v^2}\; \rho_B,\label{vvqnm}\\
V_s^q &=& g_s^q\; \sigma = {3 {g_s^q}^2\over m_s^2}\; \rho_s S,\label{vsqnm}
\end{eqnarray}
with
\begin{eqnarray}
\rho_s &=& 4 \int_0^{k_F} {d^3 k\over (2 \pi)^3}\; {M^*\over 
(k^2 + {M^*}^2)^{1/2}},\\
S &=& \int_0^R d^3 r\;\bar\psi_q \psi_q = {\Omega_q^*/2 + R m_q^*
(\Omega_q^*-1) \over \Omega_q^* (\Omega_q^*-1) + R m_q^*/2},\label{sq}
\end{eqnarray}
The last result for the quantity $S$ is an MIT bag model result\cite{st}.
The nuclear matter results are therefore determined by the solution to the
transcendental equation (\ref{vsqnm}). In terms of the redefined constants
$g_v = 3 g_v^q$ and $g_s = 3 g_s^q$, we can write
\begin{equation}
{\cal E} = 4 \int_0^{k_F} {d^3 k\over (2 \pi)^3}\; (k^2 + {M^*}^2)^{1/2}
+ {g_s^2\over 2 m_s^2} \rho_s^2 S^2 +
{g_v^2 \over 2 m_v^2} \rho_B^2.\label{etotnm}
\end{equation}

The expressions (\ref{vvqnm}), (\ref{vsqnm}), and (\ref{etotnm}) depend
only on the {\em ratio\/} of coupling constants to masses for both the
omega and sigma mesons.  Following Serot and Walecka\cite{wal}, we
introduce the two dimensionless constants
\begin{eqnarray}
C_v^2 &=& \left({g_v M \over m_v}\right)^2,\\
C_s^2 &=& \left({g_s M S\over m_s}\right)^2,
\end{eqnarray}
as free parameters.  These are adjusted to reproduce the equilibrium
properties of nuclear matter.  We note that the present model reduces to
the Walecka model in the limit that the quark mass becomes very large
while keeping the bag radius $R_0$ fixed.  This is a useful check on our
numerical codes. 

\section{Results}

In this section, we will use the MIT bag model to describe the nucleon.
The two free nuclear matter parameters $C_v^2$ and $C_s^2$ are adjusted to
reproduce a binding energy per nucleon of ${\cal E}/\rho_B - M = -15.75$
MeV at a Fermi momentum $k_F = 1.30\ {\rm fm}^{-1}$. This corresponds to a
saturation density of $\rho_B = 0.148\ {\rm fm}^{-3}$, which is the same
as that used by Serot and Walecka\cite{wal}.  A somewhat larger value of
$\rho_B = 0.17\ {\rm fm}^{-3}$ ($k_F = 1.36\ {\rm fm}^{-1}$) was used by
Saito and Thomas\cite{st}.  Our choice follows from the results of our
finite nucleus calculations.  With the larger saturation density we were
unable to simultaneously fit the rms charge radii of light and heavy
nuclei, whereas with the smaller value we could. 

Our nuclear matter results are given as models A1-A3 in Table
\ref{table2}, corresponding to bag radii $R_0$ of 0.6 fm, 0.8 fm, and 1.0
fm.  The quark mass $m_q$ has been set to 5 MeV for both $u$ and $d$
quarks.  The bag radius $R^*$ decreases slightly in-medium, although it
should be noted that the rms radius actually increases by about the same
margin.  In comparison with the results of the QHD model\cite{wal}, the
effective mass is somewhat larger and the compressibility somewhat smaller
for QMC models. Models B1 and B2 allow for a variable bag constant, and
model C uses relativistic oscillator wave functions instead of bag wave
functions.  These models are discussed in detail in the next section. 

For finite nuclei, the potentials $V_v(X)$ and $V_s(X)$ are used in the
Dirac equation (\ref{diracn}) for the nucleon to solve for the nuclear
vector and scalar densities self-consistently. Our numerical procedure is
to follow the standard iterative algorithm used in previous work\cite{hs}.
Although the in-medium form factors $v(q;X)$ and $s(q;X)$ depend on
$m_q^*(X)$, it is unnecessary to compute them at each value of $X$ and
again at each iteration.  Instead, we pre-calculate these quantities for
quark effective masses corresponding to $U_s$ ranging from 0 to 250 MeV in
steps of 1 MeV. Intermediate values of $m_q^*$ are then obtained by
interpolation. 

At this point we extend our discussion to include the $\rho$-meson and the
photon (Coulomb interaction).  The vector potentials $V_\rho(X)$ and
$V_\gamma(X)$ will have the same form as Eq.\ (\ref{vvnsph}), with the
replacement of the isoscalar vector density $\rho_v(X')$ by the
corresponding isovector density for the $\rho$-meson, and the proton
density for the photon. Because we have given the $u$ and $d$ quarks the
same mass, the quark vector form factor $v(q;X)$ is the same for the
$\rho$ and photon as for the $\omega$. 

The $\rho$-nucleon coupling constant may be taken from experiment, or from
the requirement that we reproduce the symmetry energy in nuclear matter
\cite{wal}.  Alternatively, it follows naturally within this model to
assume the universal coupling $g_\rho^q = g_v^q$, in which case
$g_\rho=g_v/3$ (since the $\rho$ couples to the isospin of the quarks). 
Because the $\omega$-nucleon coupling constant $g_v$ is much smaller here
than in QHD, all three approaches give a value of $g_\rho$ that is roughly
the same.  We choose to fix $g_\rho=2.63$, which is consistent with the
experimental value. (Note that the definition of $g_\rho$ used in
refs.\cite{st,wal} is twice as large as ours.)

As a check on our numerical codes, we were able to reproduce exactly the
results of the Walecka model\cite{wal,hs} in the limit that the quark mass
becomes very large while holding the bag radius fixed at $R^*=R_0$. 

Binding energies and rms charge radii for $^{16}$O, $^{40}$Ca, $^{90}$Zr,
and $^{208}$Pb are shown in Table \ref{table3}.  The scalar meson mass
$m_s$ has been adjusted to fit the experimental rms charge radius of
$^{40}$Ca. The binding energy per nucleon for all QMC models shows some
improvement over QHD.  We note the strong positive correlation between
binding energy per nucleon and $M^*/M$ at nuclear matter saturation
density (see Table \ref{table2}).  Charge radii are in reasonably good
agreement with experiment for all nuclei.  The charge density of $^{16}$O
for model A1 is compared with experiment in Fig.\ \ref{fig1}.  Other QMC
models and QHD give similar results. 

The improvement over QHD in the binding energies comes at the expense of a
reduction in the splitting of spin-orbit pairs, as shown in Fig.\
\ref{fig2}. This is easily understood if we recall that the spin-orbit
splitting depends on the {\em sum\/} of the vector and scalar
potentials\cite{wal}.  One can therefore expect a strong negative
correlation between spin-orbit splitting and $M^*/M$, as seen in Fig.\
\ref{fig2}.  We also note the insufficient binding of deeply bound states. 

Thus one cannot fit both the binding energy and single-particle energies
within the framework of the Hartree approximation for this class of
models. This points to the need for additional contributions to spin-orbit
effects, such as tensor correlations. 

\section{Extensions of the Model}\label{ext}
\subsection{Bag Model}

One of the possible extensions of the QMC model is to allow the bag
constants $B$ and $z$ to acquire a medium-dependence.  The parameter $z$
is supposed to incorporate center-of-mass corrections and the Casimir
effect, both of which should depend on the quark effective mass $m_q^*$. 
Initially we considered making $z$ an arbitrary linear function of $U_s$. 
However, for reasonable parameter ranges changing $z$ had little effect,
and tended to make the model worse instead of better. 

Alternatively, we can allow the bag constant $B$ to vary in the medium. 
Since quarks presumably become deconfined at high enough densities, this
would imply $B\rightarrow 0$.  This suggests a model for which $B$
decreases with increasing density.  A similar consideration has been made
in a recent preprint by Jin and Jennings\cite{jenn}, who made $B$ a
function of $M^*$. These authors also showed that for a particular
functional form of $B(M^*)$, $M^*$ becomes a linear function of the scalar
field $\sigma$.  Hence they obtained exactly the QHD model results for
nuclear matter. 

For our purposes it is more convenient to assume that $B$ is a linear
function of $U_s(X)$, such that
\begin{equation}
B^*(X) = B \left[1-\alpha_B {U_s(X)\over M}\right],
\end{equation}
with $\alpha_B$ an arbitrary parameter.  Although $B^*$ does not go to
zero at high density, we are restricting our consideration here to
densities near that of nuclear matter. Typically in nuclear matter,
$U_s \sim 130-200$ MeV for the models considered here. 

The additional dependence of $M^*(\bbox{X})$ on $U_s(\bbox{X})$, and
therefore on $\sigma(\bbox{r})$, modifies the source term of expression
(\ref{laps}) arising from Eqs.\ (\ref{lapsgen}) and (\ref{dmsigma}). 
Instead, we have
\begin{equation}
{\delta M^*(\bbox{X})\over \delta \sigma(\bbox{r})} =
-3 g_s^q \rho_s^q(\bbox{r}-\bbox{X};\bbox{X})\;\left[1 +
\case{4}{3}\pi {R^*}^3 B\;
{\alpha_B\over 3 M S(\bbox{X})}\right],
\end{equation}
which follows from Eqs.\ (\ref{diracb}), (\ref{us}), and (\ref{mstar}).
Thus the effect of this additional term can be incorporated by making the
change
\begin{equation}
s(\bbox{q};\bbox{X})\rightarrow s(\bbox{q};\bbox{X})\;
\left[1 + \case{4}{3}\pi {R^*}^3 B\;
{\alpha_B\over 3 M S(\bbox{X})}\right],
\end{equation}
to the results of section \ref{secspecific}, and a similar change to the
corresponding nuclear matter quantity $S$ appearing in
Eq.\ (\ref{vsqnm}), viz.
\begin{equation}
S\rightarrow S\; \left[1 + \case{4}{3}\pi {R^*}^3 B\;
{\alpha_B\over 3 M S}\right].
\end{equation}

The effect of $B^*$ is to reduce $M^*$ and increase the bag radius $R^*$
in-medium.  This is reflected in the results shown in Table \ref{table2}
for models labeled B1 and B2, corresponding to $\alpha_B=1$ and
$\alpha_B=2$, respectively, and $R_0=0.6$ fm. With increasing $\alpha_B$
the results and parameters of the QMC models move closer to those of the
QHD model\cite{wal}.

For finite nuclei, the results also move towards those of the QHD
model\cite{wal,hs}.  There is an improvement in spin-orbit splitting and
in the binding energy of deeply bound states, as shown in Fig.\ \ref{fig2}.
Model B2, with the largest change in the bag constant $B^*$, does
particularly well for the single particle energies and the charge density
of $^{16}$O, shown in Fig.\ \ref{fig1}.  As noted in the previous section,
an improvement in spin-orbit splitting comes at the expense of a modest
reduction in total energy (Table \ref{table3}). 

There are, however, differences between this model and QHD. All QMC models
will have electromagnetic form factors that change in the medium.  Indeed,
the charge density, or at least the quark-core contribution to the charge
density, is just the quark vector form factor $v(q;X)$ (see Appendix A).
This change arises principally from two sources.  First there is a
relative shift in the importance of the upper and lower Dirac wave
function components for quarks due to the change in effective mass $m_q^*$.
Secondly, there is a change due to the change in bag radius $R^*$.  For
models B1 and B2, the change in $R^*$ is quite large.  We will explore
the implications of medium-dependent electromagnetic form factors in a
subsequent publication.

\subsection{Relativistic Oscillator Model}

As an alternative to the bag model, we consider a relativistic oscillator
model with a (scalar) confining potential $V_{con}(r)=C r^2$, with $C=830$
MeV/fm$^2$.  This model was used by Tegen et al.\cite{tegen} to
investigate the electromagnetic properties of the quark-core in nucleons.
We refer the reader to this paper for further details. 

Our nuclear matter results given in Table \ref{table2} are quite similar
to those of bag model B1.  In particular, the increase in the rms radius
of the nucleon is of the same order as that for model B1, which in the
latter case is due predominantly to the decrease in the bag constant $B$. 
The results for the binding energies and rms charge radii are given in
Table \ref{table3}. These results are again similar to those of model B1,
as is the spin-orbit splitting shown in Fig.\ \ref{fig2}. 

The relativistic oscillator model therefore offers an alternative to the
bag models, and has one less free parameter than bag models with a
medium-dependent bag constant $B$.  Of course, we could also consider
making the oscillator constant $C$ medium-dependent.  Another advantage of
the oscillator models is that center-of-mass effects (which were
considered by Tegen et al.\cite{tegen}) are easier to handle than in bag
models. 

\section{Summary and Outlook}

In this paper we have extended previous work on Quark-Meson Coupling
models \cite{st,gu,fleck} in nuclear matter to finite nuclei.  A
generalized and systematic method for corrections to the model arising
from the spatial nonuniformity of the meson mean fields over the volume of
the nucleon is presented.  Our procedure is to define the nucleonic
potentials in terms of averages over the volume of the nucleon, as in
Eqs. (\ref{uv}) and (\ref{us}).  For this definition, the polarization
corrections are very small.

The most basic versions of this model, based on either the MIT bag
model\cite{MIT} with a fixed bag constant $B$, or a relativistic
oscillator model\cite{tegen}, have a small compressibility $K_V^{-1}$
compared with the QHD model\cite{wal}, indicating a softer equation of
state.  While this has experimental support, the drawback for these models
is that this decrease in the compressibility is accompanied by an increase
in the effective mass ratio $M^*/M$, which significantly reduces the 
spin-orbit component of the nuclear mean field and results in
insufficient binding for deeply bound states. 

One possibility we have explored is to allow the bag constant $B$ to
decrease in the medium, reducing the effective mass $M^*$ and restoring
much of the phenomenological success of the QHD model for finite
nuclei\cite{wal,hs}.  The accompanying increase in the bag radius has
implications for a number of properties of the nuclei, including
electromagnetic effects such as charge densities, form factors, and the
EMC effect\cite{emcrev}. 

In this work, we have ignored any effect of a decrease in the quark
effective mass on the masses of the virtual mesons.  Because the
parameters we use depend only on the ratio of coupling constants to
meson masses, a change in the meson mass will only result in a rescaling
of the coupling constant, without affecting the nuclear matter results.
However, there may be significant effects for finite nuclei, particularly
at the nuclear surface, and we will explore these in the future. 

We have completely neglected the center-of-mass effects for quarks in the
nucleon.  It is important to handle this problem.  In the light front
formalism, the nucleon wave function is defined so that the
intrinsic wave function is independent of the total momentum of
the nucleon. Therefore, constructing a light front version of this
model is a useful goal.

There are other possible extensions of the quark coupling model.  Aspects
of chiral symmetry have been ignored here. It should be possible to
construct a chiral version by using the cloudy bag model of the
nucleon\cite{cloudy} in which the three-quark bag is surrounded
by an evanescent cloud of pions.  Yet another aspect is that
correlations between nucleons are known to be important.  Hence the
Quark-Meson Coupling model should be extended beyond the mean
field approximation. 

The construction of a chiral version of the quark meson coupling 
model, including physics beyond the mean-field approximation,
and in which the nucleon center-of-mass degree of freedom is
treated properly, is a useful and important goal. One would have
a model consistent with conventional nuclear physics, which embodies
a reasonable treatment of the quark degrees of freedom.

\acknowledgements
One of us (PGB) would like to thank both the Institute for Nuclear Theory
and the Nuclear Theory group at the University of Washington for their
hospitality during his sabbatical leave. This work was supported in part
by the U.S.\ Department of Energy and the Natural Sciences and Engineering
Research Council of Canada. 

\newpage
\appendix
\section{Anomalous Vector Meson Coupling}

Consider the coupling of a vector meson to a quark current of the form
$j_q^\mu=g_i^q \bar\psi_q \gamma^\mu \psi_q$, governed by some generalized
quark-meson coupling constant $g_i^q$ to a quark of flavor $i$. We are
interested in computing the effective ``charge'' and ``anomalous'' form
factors governing the coupling of this meson to the composite nucleon. In
the momentum representation, the matrix elements of the quark current
between nucleon states $|N(p)>$ may be written as
\begin{equation}
\left<N(p')\left|\sum_{i=1,3} j_q^\mu(i) \right|N(p)\right> =
g\; \bar U(p') \Lambda^\mu (p'-p) U(p),
\end{equation}
where $g$ is the meson-nucleon coupling constant, $U(p)$ denotes a nucleon
spinor of four-momentum $p$, and $\Lambda^\mu$ is a vertex function of the
form
\begin{equation}
\Lambda^\mu(q) = \left[ F_1(q^2) \gamma^\mu + {F_2(q^2)\over 2M}
\imath \sigma^{\mu\nu} q_\nu \right].
\end{equation}

For quarks of equal mass, we can relate the coupling of the omega and rho
mesons to the corresponding isoscalar and isovector parts of the familiar
electromagnetic nuclear current. It is convenient to introduce the Sach's
form factors
\begin{eqnarray}
G_E(q_\mu^2) &=& F_1(q_\mu^2) - \eta F_2(q_\mu^2),\\
G_M(q_\mu^2) &=& F_1(q_\mu^2) + F_2(q_\mu^2),
\end{eqnarray}
with $\eta\equiv -q_\mu^2/4M^2$, and $q_\mu^2$ the square of the
four-momentum transfer. For individual quark wave functions of the form
\begin{equation}
\phi(\bbox{r}) = \left(\begin{array}{c}
\imath g(r)\\ \bbox{\sigma}\cdot\bbox{\hat r}
f(r)\end{array}\right)
{\chi_q \over (4 \pi)^{1/2}},
\label{quarkwf}
\end{equation}
the expressions for the proton form factors $G_{E,M}^p$ are easily derived
to leading order in $q^2\equiv |\bbox{q}|^2$: 
\begin{eqnarray}
G_E^p(q^2) &=& \int_0^\infty dr r^2\;j_0(q r) \left[g^2(r)+f^2(r)\right],\\
G_M^p(q^2) &=& -{4 M\over q} \int_0^\infty dr r^2\;j_1(q r) g(r) f(r).
\end{eqnarray}
Note that
$G_E^p(0)=\frac{1}{2}\left(G_E^{(0)}(0)+G_E^{(1)}(0)\right)=1$ and 
$G_E^n(0)=\frac{1}{2}\left(G_E^{(0)}(0)-G_E^{(1)}(0)\right)=0$, so that
$G_E^{(0,1)}(0)=F_1^{(0,1)}(0)=1$
for both the isoscalar and isovector form factors. The magnetic moment of
the proton due to the quark core is given by
\begin{equation}
\mu_p = {e\over 2M}\;G_M^p(0) =
-{2e\over 3} \int_0^\infty dr r^3\; g(r) f(r),
\end{equation}
and the standard quark model ratio $\mu_n = -\frac{2}{3}\mu_p$ gives the
magnetic moment of the neutron.
The form factor $G_E^{(0)}(q)$ is just the quark vector form factor $v(q)$
defined in Eq.\ (\ref{vform}).  We expect $g_\omega=3 g_\omega^q$ for the
(isoscalar) $\omega$-nucleon coupling constant, and $g_\rho = g_\rho^q$ for
the (isovector) $\rho$-nucleon coupling constant.

Experimentally, $F_2^{(0)}(0)=-.1203$ and $F_2^{(1)}(0)=3.706$ for the
electromagnetic form factors. Although this includes contributions besides
the quark core, such as the pionic cloud, we expect the ``anomalous''
coupling of the vector mesons to the quark core to be of the same order
as $F_2^{(0,1)}$.  Because the isoscalar anomalous coupling term is so
small, we have chosen to ignore it in the calculations presented here. 

For completeness, we give the appropriate modifications of our formalism
when the anomalous tensor terms are included.  With the introduction of a
``tensor'' potential $V_T(\bbox{X})$, the Dirac equation (\ref{diracn})
for the nucleon becomes
\begin{equation}
\left[-\imath\bbox{\gamma}\cdot\bbox{\nabla}
+ M^*(\bbox{X})+\gamma^0 V_v(\bbox{X})
+\imath\gamma^0 \bbox{\gamma}\cdot\hat{\bbox{X}} V_T(\bbox{X})
\right]\psi_i(\bbox{X})=
E_i \gamma^0 \psi_i(\bbox{X}).
\end{equation}
It is convenient to introduce the ``tensor'' density
\begin{equation}
\rho_T(\bbox{X})=\sum_{i=occ}\imath\bar\psi_i(\bbox{X})
\gamma^0 \bbox{\gamma}\cdot\hat{\bbox{X}}
\psi_i(\bbox{X}).
\end{equation}
Assuming the nucleon wave functions are defined analogously to
Eq.\ (\ref{quarkwf}), we have $\rho_T(X)= 2 \sum_i G_i(X) F_i(X)$.

For spherical nuclei, the vector potential $V_v(X)$ of Eq.\ (\ref{vvnsph})
becomes
\begin{eqnarray}
V_v(X) &=& g_v^2 {2 \over \pi} \int_0^\infty
dq\; {q^2\over q^2+m_v^2}\; j_0(q X) F_1(q;X)
\int_0^\infty dX'\; X'^2\; \Biggl[ j_0(q X') \rho_v(X')
F_1(q;X')\nonumber\\
&& +\; q j_1(q X') \rho_T(X') {F_2(q;X')\over 2M}\Biggr],
\end{eqnarray}
and the potential $V_T(X)$ is
\begin{eqnarray}
V_T(X) &=& g_v^2 {2 \over \pi} \int_0^\infty
dq\; {q^3\over q^2+m_v^2}\;j_1(q X){F_2(q;X)\over 2M}
\int_0^\infty dX'\; X'^2\; \Biggl[ j_0(q X') \rho_v(X')
F_1(q;X')\nonumber\\
&& +\; q j_1(q X') \rho_T(X') {F_2(q;X')\over 2M}\Biggr].
\end{eqnarray}
Note that $V_T(X)\propto-d V_v(X)/dX$ in the absence of a medium-dependence
to the form factors. Finally, the contribution of the vector meson field
to the total energy is modified from Eq.\ (\ref{eomega2}) to
\begin{equation}
E_v = - \case{1}{2} \int d^3 X\;\left[\rho_v(\bbox{X}) V_v(\bbox{X})
+ \rho_T(\bbox{X}) V_T(\bbox{X})\right],
\end{equation}
which again removes half of the contribution of the vector potential
contained in the nucleon energies.

\section{Polarization Corrections}

Here we consider the second order perturbative corrections to the vector
potential $U_v(\bbox{X})$ and the quark energy $\epsilon^*(\bbox{X})$
arising from the residual interaction $\Delta U_{v,s}(\bbox{r},\bbox{X})$. 
We also examine the first order corrections to matrix elements. All of
these corrections involve intermediate excited states of the nucleon,
hence the title of this Appendix. 

We study the corrections involving only the vector interaction here, and
show that polarization effects are vanishingly small. The formalism and
physics of the scalar terms is similar, so we expect the same negligible
effects. 

It is useful to introduce the braket notation, so that
\begin{equation}
<\bbox{r}|\phi(\bbox{X})>\equiv \phi(\bbox{r}-\bbox{X},\bbox{X}),
\end{equation}
in the coordinate space representation. With this notation, the residual
interaction can be expressed as (see Eqs.\ (\ref{udef}) and (\ref{us}))
\begin{equation}
\Delta U_v(\bbox{r},\bbox{X}) \equiv V_v^q(\bbox{r}) -
<\phi(\bbox{X})|V_v^q(\bbox{r})|\phi(\bbox{X})>,
\end{equation}
so that the first order perturbation theory correction to $U_v(\bbox{X})$
vanishes:
\begin{equation}
\delta U_v^{(1)}(\bbox{X}) =
<\phi(\bbox{X})|\Delta U_v(\bbox{r},\bbox{X})|\phi(\bbox{X})>=0.
\end{equation}

The second order correction is given from standard Rayleigh-Schroedinger
perturbation theory as
\begin{equation}
\delta U_v^{(2)}(\bbox{X})=\sum_{n\ne 
N}{\left|<N(\bbox{X})|\sum_{i=1,3}\Delta 
U_v(\bbox{r}_i,\bbox{X})|n(\bbox{X})>\right|^2\over 
E_N(\bbox{X})-E_n(\bbox{X})},
\label{second}
\end{equation}
where $N$ and $n$ represent the nucleon and excited baryon states,
respectively. The sum on $i$ is over the three quarks in the baryon. The
confinement of the quarks to the region near $\bbox{X}$ enables a Taylor
series expansion of $\Delta U_v(\bbox{r}_i,\bbox{X})$ about the point
$\bbox{r}_i=\bbox{X}$.  Let us define a variable
\begin{equation}
\bbox{s}^{(i)}\equiv \bbox{r}_i-\bbox{X},
\end{equation}
so that 
\begin{eqnarray}
\Delta U_v(\bbox{s}^{(i)},\bbox{X})\approx
V_v^q(\bbox{X}) &-&
<\phi(\bbox{X})|V_v^q(\bbox{r})|\phi(\bbox{X})>
+\bbox{s}^{(i)}\cdot\bbox{\nabla}V_v^q(\bbox{X})\nonumber\\
&+&\case{1}{2}\sum_{\alpha,\beta=1,3}s^{(i)}_\alpha s^{(i)}_\beta
{\partial^2 V^q_v\over \partial X_\alpha\partial X_\beta}(\bbox{X}).
\label{expand}
\end{eqnarray}
The first three terms of Eq.\ (\ref{expand}) do not lead to an excitation
of the nucleon and do not contribute to the sum on $n$ of Eq.\
(\ref{second}). This is because the first two terms do not contain any
dependence on $\bbox{s}^{(i)}$, and the third term is a center-of-mass
operator. The last term of Eq.\ (\ref{expand}) can be simplified by
regrouping the term $s^{(i)}_\alpha s^{(i)}_\beta$ into quadrupole and
monopole terms: 
\begin{eqnarray}
s^{(i)}_\alpha s^{(i)}_\beta &=& s^{(i)}_\alpha s^{(i)}_\beta 
-\delta_{\alpha,\beta}\case{1}{3}\bbox{s}^{(i)}\cdot \bbox{s}^{(i)}
+\delta_{\alpha,\beta}\case{1}{3}\bbox{s}^{(i)}\cdot \bbox{s}^{(i)}
\nonumber\\ &\equiv& Q^{(i)}_{\alpha\beta}
 +\delta_{\alpha,\beta}\case{1}{3}\bbox{s}^{(i)}\cdot \bbox{s}^{(i)}.
\label{q2s}
\end{eqnarray}
The quadrupole term, when combined with the spherically symmetric nature
of $V_v^q(X)$, leads to a quadrupole dependence on the nuclear
coordinates. The average of this vanishes for our spherical shell model.
Using the monopole term of Eq.\ (\ref{q2s}) in Eqs.\ (\ref{expand}) and
(\ref{second}) leads to the result
\begin{equation}
\delta U_v^{(2)}(\bbox{X})=\sum_{n\ne 
N}{\left|<N(\bbox{X})|\sum_{i=1,3}\case{1}{6}\bbox{s}^{(i)}
\cdot\bbox{s}^{(i)}|n(\bbox{X})>\right|^2\over 
E_N(\bbox{X})-E_n(\bbox{X})}\left(\nabla^2U_v(\bbox{X})\right)^2.
\label{secondr}
\end{equation}
This is an attractive second order contribution.

We shall proceed to obtain a numerical estimate of the size of the
correction given by Eq.\ (\ref{secondr}). First we shall assume that the
isoscalar monopole excitation strength is taken up by a single state ---
the Roper resonance. This allows us to replace the different values of
$E_N(\bbox{X})-E_n(\bbox{X})$ by a single one, $\Delta E_R\approx -500$
MeV. The sum over $n$ can then be performed using closure, so that
\begin{equation}
\delta U_v^{(2)}(\bbox{X})\approx {1\over 36} 
{\nabla^2 U_v(\bbox{X})^2\over \Delta E}<N(\bbox{X})|\left(
\rho^2-<\rho^2>+\lambda^2 -<\lambda^2 >\right)^2|N(\bbox{X})>,
\end{equation}
in which the Jacobi coordinates 
$\bbox{\rho}\equiv{1\over\sqrt{2}}(\bbox{s}^{(1)}-\bbox{s}^{(2)})$ and
$\bbox{\lambda}\equiv {1\over \sqrt{6}}
(\bbox{s}^{(1)}+\bbox{s}^{(2)}-2\bbox{s}^{(3)})$
are used, and $<\rho^2(\lambda^2)>$ represents the nucleonic matrix
element of $\rho^2$ ($\lambda^2$). Simple algebra, and the reasonable
assumption that $<\lambda^2>=<\rho^2>$, which holds exactly in the
hyperspherical approximation, leads to the expression
\begin{equation}
\delta U_v^{(2)}(\bbox{X})\approx {1\over 18} 
{\nabla^2U_v(\bbox{X})^2\over \Delta E}\;\left(
<\rho^4>-<\rho^2>^2\right).
\end{equation}

Dimensional analysis yields the result that $<\rho^4>-<\rho^2>^2\propto
r_N^4$, where $r_N$ is the root mean square radius of the nucleon. We
proceed by using a simple Woods-Saxon form for $U_v(\bbox{X})$,
\begin{equation}
U_v(\bbox{X})= V_0 / \left[1+e^{(X-R_A)/a}\right].
\end{equation}
Typically, $R_A=r_0\;A^{1/3}$, with $r_0\approx 1.1-1.2$ and $a=0.54$ fm.

The density is approximately constant near the center of the nucleus, so
that $\nabla^2 U_v(X)$ can be effective only near the nuclear surface. For
example, at $X=R_A$
\begin{equation}
\delta U_v^{(2)}(R_A)\approx {1\over 18} 
{V_0^2\over(2 a R_A)^2 \Delta E}\;\left(
<\rho^4>-<\rho^2>^2\right).
\end{equation}
If one takes $a$ and $r_N$ to each be of the order of the nucleon
size, one sees that $\delta U_v^{(2)}(R_A)$ is proportional to
$(r_N/R_A)^2$, which is very small, a result that has been obtained
without using a specific model of confinement. One may get a specific
estimate by using the non-relativistic harmonic oscillator quark model of
the nucleon. In this case, the closure approximation used above is true
exactly, and $<\rho^4>-<\rho^2>^2=1.5\;r_N^4$. Then, using typical values
$r_N=0.83$ fm, $r_0=1.15$ fm, $a=0.54$ fm, and $V_0=250$ MeV, one obtains
the result
\begin{equation}
\delta U_v^{(2)}(R_A)\approx -{3\over A^{2/3}}\ {\rm MeV},
\end{equation}
which ranges from $-0.5$ MeV for $^{16}$O to $-0.1$ MeV for $^{208}$Pb.
These effects are truly small. 

We next consider the effects of $\rho$ meson exchange. The equation that is
analogous to Eq.\ (\ref{expand}) is 
\begin{equation}
\delta U_\rho^{(2)}(\bbox{s}^{(i)},\bbox{X})\approx
V_\rho^q(\bbox{X}) -
<\phi(\bbox{X})|V_\rho^q(\bbox{r})|\phi(\bbox{X})>
+\tau_3^{(i)}\;\bbox{s}^{(i)}\cdot\bbox{\nabla} V_v^q(\bbox{X}),
\end{equation}
in which the term $\sum_i\tau_3^{(i)}\bbox{s}^{(i)}$ is the dipole
operator $\bbox{D}^{(i)}$. Now the term linear in $\bbox{s}^{(i)}$ does
contribute.  We thus obtain the second order correction to the $\rho$
exchange contribution to the vector potential:
\begin{equation}
\delta U_\rho^{(2)}(\bbox{X})=\sum_{n\ne N}{
\left|<N(\bbox{X})|\sum_{i=1,3}
\bbox{D}^{(i)}|n(\bbox{X})>\cdot\bbox{\nabla}U_\rho(\bbox{X})
\right|^2\over E_N(\bbox{X})-E_n(\bbox{X})}.
\end{equation}
This term can be related to the experimentally measured dipole
polarizability $\alpha$, which is given by
\begin{equation}
\alpha\equiv 2 e^2\sum_{n\ne N}{
\left|<N(\bbox{X})|\sum_{i=1,3}
D_3^{(i)}|n(\bbox{X})>\right|^2 \over
E_N(\bbox{X})-E_n(\bbox{X})}.
\end{equation}
The use of the Wigner-Eckart theorem allows one to obtain the relation
between $\delta U_\rho^{(2)}$ and the measured quantity $\alpha$:
\begin{equation}
\delta U_\rho^{(2)}(\bbox{X})=-{|\alpha|\over e^2}\;
\left|\bbox{\nabla}U_\rho(\bbox{X})\right|^2
\end{equation}

The use of the value $\alpha=12\times 10^{-4}$ fm$^3$ (as explained in the
review \cite{lvov}) and the Woods-Saxon form of the potential, along with
the coupling constants discussed above, leads to the estimate $\delta
U_\rho^{(2)}(X=R_A)=-0.2 $ MeV, so that this effect is negligible. 

The previous examples show that the second order contributions to the
vector potentials are very small. The use of the same formalism leads to
the result that the second order contributions to the scalar potential are
also negligible. 

Our final example concerns the computations of matrix elements.  Consider
some general observable ${\cal O}$. The first order correction supplied by
the vector potential is given by
\begin{equation}
\delta {\cal O}_v=2\sum_{n\ne N}{<N(\bbox{X})|{\cal 
O}|n(\bbox{X})><n(\bbox{X})|\Delta U_v|N(\bbox{X})>\over
E_N(\bbox{X})-E_n(\bbox{X})}.
\end{equation}
To be specific, consider the case ${\cal O}=\sum_i \bbox{s}^{(i)}\cdot
\bbox{s}^{(i)}\equiv s^2$. In this case we obtain
\begin{equation}
\delta s^2_v(X)=\case{1}{3}\sum_{n\ne N}{\left|<N(X)|s^2|n(X)>
\right|^2\over E_N(X)-E_n(X)}\;\nabla^2U_v(X),
\end{equation}
or, using the approximations used to obtain the corrections to the energy,
\begin{equation}
\delta s^2_v(X)={2\over 3}{<\rho^4>-<\rho^2>^2\over
\Delta E}\;\nabla^2 U_v(X).
\end{equation}
In this case the contribution of the attractive scalar potential tends to
cancel that of the repulsive vector potential.  We account for this in a
simple way by taking a net potential of depth 50 MeV. The net result is
that
\begin{equation}
{\delta s^2(X=R_A)\over <s^2>}\approx -{0.01\over A^{1/3}},
\end{equation}
which is again a negligible effect. 

The final result is that the dominant effects of the composite nature of
the nucleon are contained in the definitions (\ref{uv}) and (\ref{us}) of
the mean quark vector and scalar potentials as expectation values of the
quark vector and scalar potentials in the nucleon.

\narrowtext
\begin{table}
\caption{Nucleon bag model parameters corresponding to $m_q = 5$ MeV and
a free nucleon mass of $M=938.9$ MeV.}
\begin{tabular}{lddd}
$R_0$ (fm)&$\case{4}{3}\pi R_0^3 B$ (MeV)&$z$\\
\tableline
0.6 & 232.918 & 4.00367\\
0.8 & 232.915 & 3.29545\\
1.0 & 232.912 & 2.58724\\
\end{tabular}
\label{table1}
\end{table}

\begin{table}
\caption{Model parameters corresponding to a nuclear matter
binding energy per nucleon of $-15.75$ MeV at $k_F =$ 1.30 fm$^{-1}$.
Nuclear matter results are given for the compressibility, nuclear
effective mass, and in-medium bag radius (rms radius for the oscillator
model).}
\begin{tabular}{ldddddd}
Model&$R_0$ (fm)&$C_s^2$&$C_v^2$&$K_V^{-1}$ (MeV)&$M^*/M$&$R^*$ (fm)\\
%&(fm)&&&(MeV)&&(fm)\\
\tableline
QHD &     & 357.62 & 274.00 & 547 & 0.541 &\\
A1  & 0.6 & 154.94 & 117.46 & 292 & 0.775 & 0.596\\
A2  & 0.8 & 132.93 &  98.41 & 279 & 0.803 & 0.795\\
A3  & 1.0 & 117.15 &  84.57 & 266 & 0.823 & 0.993\\
B1 ($\alpha_B=1$) & 0.6 & 187.98 & 145.50 & 321 & 0.734 & 0.624\\
B2 ($\alpha_B=2$) & 0.6 & 289.53 & 226.11 & 364 & 0.614 & 0.676\\
C (Rel.\ Osc.) & 0.642\tablenotemark[1] & 190.09 & 147.26 & 326 & 0.731 &
0.689\tablenotemark[1]\\
\end{tabular}
\tablenotetext[1]{For the relativistic oscillator we give the rms radius.}
\label{table2}
\end{table}

\begin{table}
\caption{Binding energy per nucleon (in MeV), and rms charge radii (in fm)
for several closed shell nuclei.  We have taken $m_v=783$ MeV,
$m_\rho=770$ MeV, and $g_\rho=2.63$.  The scalar meson mass $m_s$ has been
adjusted to fit the rms charge radius of 3.48 fm in $^{40}$Ca.}
\begin{tabular}{lddddddddddd}
Model&$m_s$&\multicolumn{2}{c}{$^{16}$O}&\multicolumn{2}{c}{$^{40}$Ca}&
\multicolumn{2}{c}{$^{90}$Zr}&\multicolumn{2}{c}{$^{208}$Pb}\\
\cline{3-4} \cline{5-6} \cline{7-8} \cline{9-10}
&(MeV)&${\cal E}/A$&$<r^2>^{1/2}_{ch}$&${\cal E}/A$&$<r^2>^{1/2}_{ch}$&
${\cal E}/A$&$<r^2>^{1/2}_{ch}$&${\cal E}/A$&$<r^2>^{1/2}_{ch}$\\
\tableline
QHD & 520 & -4.88 & 2.74 & -6.30 & 3.48\tablenotemark[1] & -7.07 & 4.27 &
-6.82 & 5.48\\
A1  & 450 & -6.22 & 2.75 & -7.46 & 3.48\tablenotemark[1] & -7.84 & 4.32 &
-7.59 & 5.59\\
A2  & 415 & -5.98 & 2.75 & -7.30 & 3.48\tablenotemark[1] & -7.70 & 4.32 &
-7.49 & 5.59\\
A3  & 385 & -5.72 & 2.75 & -7.12 & 3.48\tablenotemark[1] & -7.56 & 4.32 &
-7.40 & 5.59\\
B1  & 460 & -5.93 & 2.75 & -7.23 & 3.48\tablenotemark[1] & -7.68 & 4.31 &
-7.46 & 5.57\\
B2  & 480 & -5.57 & 2.75 & -6.94 & 3.48\tablenotemark[1] & -7.55 & 4.29 &
-7.34 & 5.52\\
C   & 455 & -5.51 & 2.75 & -6.94 & 3.48\tablenotemark[1] & -7.46 & 4.31 &
-7.30 & 5.56\\
Expt & & -7.98 & 2.75 & -8.55 & 3.48 & -8.71  & 4.28 & -7.87 & 5.50\\
\end{tabular}
\tablenotetext[1]{Fit}
\label{table3}
\end{table}

\begin{figure}
\caption{The charge density of $^{16}$O for bag models A1 and B2, which
both have $R_0=0.6$ fm.  Model B2 has a medium-dependent bag constant. 
The solid line is the experimental parameterization of
H.\ de Vries, C.\ W.\ de Jager and C.\ de Vries, At.\ 
Data Nucl.\ Data Tables {\bf 36}, 495 (1987).}
\label{fig1}
\end{figure}

\begin{figure}
\caption{Proton single particle energies in $^{16}$O for all models.}
\label{fig2}
\end{figure}

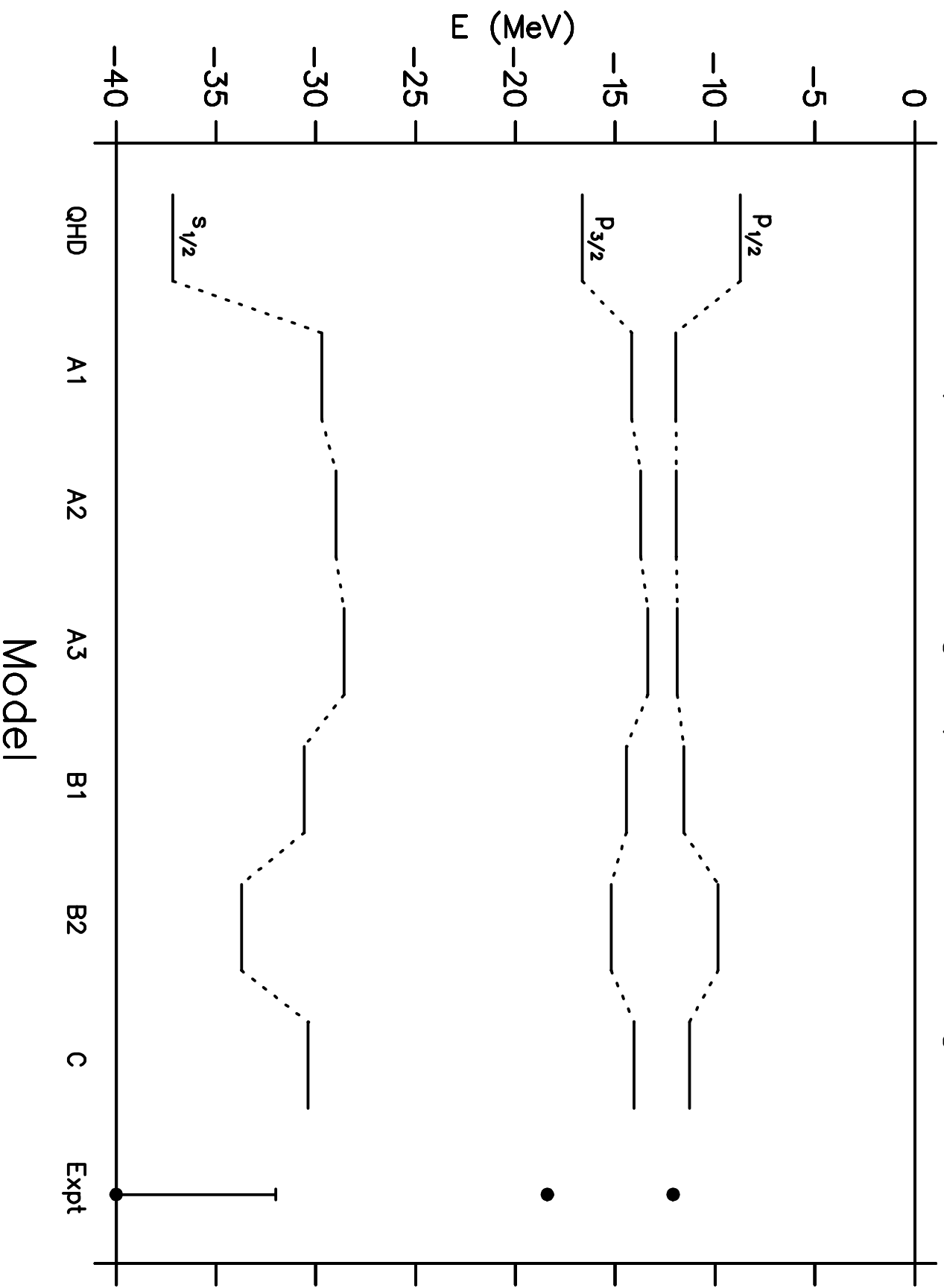
\end{document}